\documentclass[preprint2]{emulateapj}
\usepackage{graphicx}
\usepackage{graphicx}
\usepackage{txfonts}
\usepackage{url}
\usepackage{bm}
\graphicspath{{./fig/}{./png/}}

\newcommand{\EQ}{\begin{equation}}
\newcommand{\EE}{\end{equation}}
\newcommand{\EN}{\end{equation}}
\newcommand{\EQA}{\begin{eqnarray}}
\newcommand{\EEA}{\end{eqnarray}}
\newcommand{\brac}[1]{\langle #1 \rangle}

\newcommand{\EEq}[1]{Equation~(\ref{#1})}
\newcommand{\Fig}[1]{Fig.~\ref{#1}}
\newcommand{\FFig}[1]{Figure~\ref{#1}} 
\newcommand{\bra}[1]{\langle{#1}\rangle}
\newcommand{\mean}[1]{\overline{#1}}
\newcommand{\meanv}[1]{\overline{\bm #1}}

\newcommand{\urms}{u_{\rm rms}}
\newcommand{\brms}{B_{\rm rms}}

\newcommand{\xu}{\hat{\bm x}}
\newcommand{\yu}{\hat{\bm y}}

\newcommand{\kef}{k_{\rm f}}

\newcommand{\St}{{\rm St}}

\def\Pm{\mbox{\rm Pr}_M}
\def\Rm{\mbox{\rm Re}_M}
\def\Rey{\mbox{\rm Re}}
\newcommand{\Sh}{{\rm Sh}}
\newcommand{\kf}{k_{\rm f}}
\newcommand{\epsf}{\epsilon_{\rm f}}
\newcommand{\epsm}{\epsilon_{\rm m}}
\newcommand{\Beq}{B_{\rm eq}}

\newcommand{\etat}{\eta_{\rm t}}
\newcommand{\etatz}{\eta_{\rm t0}}
\newcommand{\etaT}{\eta_{\rm T}}
\newcommand{\meanAA}{\overline{\bm A}}
\newcommand{\meanBB}{\overline{\bm B}}
\newcommand{\meanJJ}{\overline{\bm J}}

\newcommand{\meanB}{\overline{B}}
\newcommand{\eee}{\hat{\bm e}}
\newcommand{\bb}{\bm b}
\newcommand{\ff}{\bm f}
\newcommand{\jj}{\bm j}
\newcommand{\kk}{\bm k}
\newcommand{\xx}{\bm x}
\newcommand{\aaa}{\bm a}
\newcommand{\AAA}{\bm A}
\newcommand{\BB}{\bm B}
\newcommand{\FF}{\bm F}
\newcommand{\JJ}{\bm J}
\newcommand{\UU}{\bm U}
\newcommand{\nab}{\bm\nabla}
\newcommand{\ii}{{\rm i}}
\newcommand{\const}{{\rm const}}
\newcommand{\RRRR}{\mathsf R}
\newcommand{\DDD}{{\cal D}}

\begin{document}
\title{Turbulent dynamos with shear and fractional helicity}
\author{Petri J.\ K\"apyl\"a\altaffilmark{1} and Axel Brandenburg\altaffilmark{2}}

\altaffiltext{1}{
Observatory, University of Helsinki, PO Box 14, FI-00014 University of Helsinki, Finland
}\altaffiltext{2}{
NORDITA, Roslagstullsbacken 23, SE-10691 Stockholm, Sweden
\\ $ $Revision: 1.101 $ $ (\today)
}

\begin{abstract}
  Dynamo action owing to helically forced turbulence and large-scale
  shear is studied using direct numerical simulations.
  The resulting magnetic field displays propagating wave-like behavior.
  This behavior can be modelled in terms of an $\alpha\Omega$ dynamo.
  In most cases
  super-equipartition fields are generated. By varying the
  fraction of helicity of the turbulence the regeneration of poloidal fields
  via the helicity effect (corresponding to the $\alpha$-effect) is regulated.
  The saturation level of the magnetic field in the numerical models
  is consistent with a linear dependence on the ratio of the fractional helicities 
  of the small and large-scale fields, as predicted by a simple nonlinear 
  mean-field model.
  As the magnetic Reynolds number ($\Rm$) based on the wavenumber of the
  energy-carrying eddies is increased from 1 to 180, the cycle frequency of the
  large-scale field is found to decrease by a factor of about 6
  in cases where the turbulence is fully helical.
  This is interpreted in terms of the turbulent magnetic diffusivity,
  which is found to be only weakly dependent on $\Rm$.
\end{abstract}


\section{Introduction}

Several classes of turbulent astrophysical bodies including stars with
outer convection zones and spiral galaxies can contain pronounced 
large-scale magnetic fields.
The scales of these fields are much larger than the scale of the
energy-containing eddies of the turbulence responsible for producing
these fields.
Explaining such fields is an important aim of dynamo theory.
Today, turbulence simulations begin to reproduce the phenomenon
of large-scale field generation
(e.g.\ Brandenburg 2001; Brandenburg et al.\ 2001, 2008a;
Yousef et al.\ 2008a,b; K\"apyl\"a et al.\ 2008, 2009a; Hughes \& Proctor 2009)
and allow making contact with
mean-field dynamo theory, which parameterizes the effects of
small-scale correlations such as kinetic helicity on the evolution of the
large-scale field (Moffatt 1978; Parker 1979; Krause \& R\"adler 1980).
Indeed, helicity has long been known to facilitate the production of large-scale
fields under the condition that an appropriately defined dynamo number
exceeds a critical value.

Significant progress in the nonlinear formulation of large-scale dynamo
theory has been possible by carefully designing and comparing simulations
of turbulent magnetic fields with mean-field theory.
It is important that both are
applied to equivalent situations that are sufficiently simple (simple geometries
and boundary conditions, homogeneity of the turbulence, etc).
Particularly illuminating is the case with triply-periodic boundary conditions
where the energy of the large-scale field can exceed the energy of
the small-scale field by a factor that is equal to the scale separation
ratio, i.e.\ the scale of the system divided by the scale of the
energy-carrying eddies.
The phenomenology of this behavior based on magnetic helicity conservation
was already described by Brandenburg \cite{Brandenburg2001}.
He found that full saturation occurs on a resistive time scale following
a characteristic einschalt or switch-on curve pattern.
The corresponding nonlinear mean-field theory for helical dynamos was
developed by Field \& Blackman \cite{FieldBlack2002}, and extended to the case
with shear by Blackman \& Brandenburg (2002, hereafter BB02).

One of the important and surprising findings since the early works of
Cattaneo \& Vainshtein (1991), Vainshtein \& Cattaneo (1992),
Gruzinov \& Diamond (1994), and Bhattacharjee \& Yuan (1995)
was the realization that the value of the magnetic Reynolds number, $\Rm$,
enters the nonlinear mean-field theory.
Even today simulations have not yet been able to establish rigorously
that the large-scale dynamo solutions obtained so far are asymptotically independent
of the value of $\Rm$.
There are many aspects of this problem;
the most important one is probably the possibility of catastrophic
$\alpha$ quenching.
This means that the field cannot saturate to equipartition strength
on a dynamical timescale and that in a closed or periodic domain fields
of equipartition strength can only be reached on a resistive timescale.
It is fairly clear now that catastrophic quenching can only
be alleviated in the presence of magnetic helicity fluxes
(Blackman \& Field 2000; Kleeorin et al.\ 2000).
What is less clear, however, is whether the helicity fluxes themselves depend on
$\Rm$ and on the mean field.

Numerical evidence for large-scale fields in the presence of boundaries
comes from simulations of both forced turbulence (Brandenburg 2005)
as well as convection (K\"apyl\"a et al.\ 2008).
The latter reference was particularly effective in explaining the reason
for the absence of a significant large-scale field in the simulations of
Tobias et al.\ (2008), even though their boundary conditions would have
allowed a helicity flux.
The reason is that the helicity flux follows the direction of the contours
of constant shear (Brandenburg \& Subramanian 2005b), but in the simulations
of Tobias et al.\ (2008) these contours do not cross an open surface,
because they used periodic boundary conditions in the lateral direction.
When using instead open boundary conditions in the lateral direction, a
strong large scale field is obtained (see Figure~17 in K\"apyl\"a et al.\ 2008).
Alternatively, one can use vertical contours of constant shear, as was done
in K\"apyl\"a et al.\ (2008).
This was recently confirmed by Hughes \& Proctor (2009) in an independent
study.

In the absence of magnetic helicity fluxes and without boundaries,
strong large-scale fields can only be generated on a resistive timescale
(Brandenburg 2001; Brandenburg et al.\ 2001).
With boundaries, and in the absence of magnetic helicity fluxes,
the strength of large-scale magnetic fields decreases like $\Rm^{-1/2}$
with increasing values of $\Rm$; see Brandenburg \& Subramanian (2005b).
Recent simulations of rigidly rotating convection have also been
successful in generating a large-scale dynamo (K\"apyl\"a et al.\
2009a). There, however, the $\Rm$-coverage is not sufficient to
determine a scaling for the saturation level.
In the following we focus on the former case using periodic boundary
conditions, so no magnetic helicity can leave the domain, but large-scale
fields can still emerge on a resistive timescale.
We use such a configuration to study the quenching of the
turbulent magnetic diffusivity, $\etat$, and whether its value depends
on $\Rm$.
The quenched value of $\etat$ is crucial for determining the cycle
period in an oscillatory $\alpha\Omega$ dynamo.
Indeed, BB02 showed that in the saturated regime
and under the assumption of homogeneity the
cycle frequency $\omega_{\rm cyc}$ of an $\alpha \Omega$ dynamo can be
written as
\EQ
\omega_{\rm cyc} = \etaT k_{\rm m}^2,
\label{ocyc}
\EE
where $\etaT = \etat + \eta$ is the total (turbulent plus microscopic)
magnetic diffusivity in the quenched state
and $k_{\rm m}$ is the wavenumber of the mean magnetic field.
Experiments with nonlinear mean-field models suggest that Equation~(\ref{ocyc})
remains approximately valid even in the mildly nonlinear regime.

\EEq{ocyc} applies to homogeneous dynamo waves of sufficiently
low amplitude, so that the field variation remains harmonic in
space and time.
Such behavior can only be modeled with periodic boundary conditions
in the direction perpendicular to the plane of the shear flow.
The result is expected to change in the presence of boundary conditions,
but $\omega_{\rm cyc}$ should still be proportional to $\etaT$.
Therefore, any $\Rm$-dependent quenching of $\etaT$ would directly
affect $\omega_{\rm cyc}$, regardless of whether or not magnetic
helicity fluxes would alleviate catastrophic $\alpha$ quenching.
This underlines the importance of studying the quenching of $\etat$.

Determining the quenching of $\etat$ has been attempted on various
occasions in the past.
It is now generally accepted that, in two dimensions, $\etat$ is
catastrophically quenched proportional to $1/\Rm$, provided
the energy of the mean field is comparable to the energy of the
turbulence.
However, this is because in two dimensions the mean squared magnetic
vector potential is conserved, which would not be applicable in
three dimensions (Gruzinov \& Diamond 1995).
For three-dimensional turbulence, analytic theory predicts
$\etat$ quenching proportional to the strength of the mean field
independently of $\Rm$.
Such a behavior also emerges from forced turbulence simulations
where an initially sinusoidal magnetic field is found to decay proportional
to $\exp(-\etaT k_{\rm m}^2 t)$ (Yousef et al.\ 2003).
The inferred quenching formula, applied to the corresponding
mean-field dynamo models, is also found to produce the best fit 
to simulations that do show strong large-scale dynamo action (BB02).
More recently, Brandenburg et al.\ (2008b) have computed the full
$\alpha_{ij}$ and $\eta_{ij}$ tensors in the saturated state
without shear, where the mean field is a Beltrami field.
In that case, the turbulent magnetic diffusivity was found to be
reduced by a factor of about 5 as $\Rm$ is increased from 2 to 600.
However, the case of Beltrami fields has certain limitations that are
avoided when there is shear.

\section{The simulations}
\label{ForcingFunction}

We solve the stochastically forced isothermal hydromagnetic equations
in a cubical domain of size $(2\pi)^3$ in the
presence of a uniform shear flow, $\UU_0=(0,Sx,0)$, with $S=\const$,
\EQ
{\DDD\AAA\over\DDD t}=-S A_y\xu-(\nab\UU)^T\AAA+\eta\nabla^2\AAA,
\EN
\EQ
{\DDD\UU\over\DDD t}=-S U_x\yu-c_{\rm s}^2\nab\ln\rho
+{1\over\rho}\JJ\times\BB+\FF_{\rm visc}+\ff,
\EN
\EQ
{\DDD\ln\rho\over\DDD t}=-\nab\cdot\UU,
\EN
where $\DDD/\DDD t=\partial/\partial t+(\UU+\UU_0)\cdot\nab$ is the
advective derivative with respect to the total flow velocity that
also includes the shear flow, and $[(\nab\UU)^T\AAA]_i=U_jA_{i,j}$
in component form, $\FF_{\rm visc}$ is the viscous force,
and $\ff$ is the forcing term.
As in earlier work (Brandenburg 2001),
the forcing function is given by
\EQ
\ff(\xx,t)={\rm Re}\{N\ff_{\kk(t)}\exp[\ii\kk(t)\cdot\xx+\ii\phi(t)]\},
\EN
where $\xx$ is the position vector.
The wavevector $\kk(t)$ and the random phase
$-\pi<\phi(t)\le\pi$ change at every time step, so $\ff(\xx,t)$ is
$\delta$-correlated in time.
For the time-integrated forcing function to be independent
of the length of the time step $\delta t$, the normalization factor $N$
has to be proportional to $\delta t^{-1/2}$.
On dimensional grounds it is chosen to be
$N=f_0 c_{\rm s}(|\kk|c_{\rm s}/\delta t)^{1/2}$, where $f_0$ is a
nondimensional forcing amplitude.
At each timestep we select randomly one of many possible wavevectors
in a certain range around a given forcing wavenumber.
The average wavenumber is referred to as $k_{\rm f}$.
We force the system with transverse helical waves,
\begin{equation}
\ff_{\kk}=\RRRR\cdot\ff_{\kk}^{\rm(nohel)}\quad\mbox{with}\quad
{\sf R}_{ij}={\delta_{ij}-\ii\sigma\epsilon_{ijk}\hat{k}_k
\over\sqrt{1+\sigma^2}},
\end{equation}
where $\sigma=1$ for the fully helical case with
positive helicity of the forcing function,
\EQ
\ff_{\kk}^{\rm(nohel)}=
\left(\kk\times\eee\right)/\sqrt{\kk^2-(\kk\cdot\eee)^2},
\label{nohel_forcing}
\EN
is a non-helical forcing function, and $\eee$ is an arbitrary unit vector
not aligned with $\kk$; note that $|\ff_{\kk}|^2=1$.

\begin{figure*}[t]
\centering\includegraphics[width=0.8\textwidth]{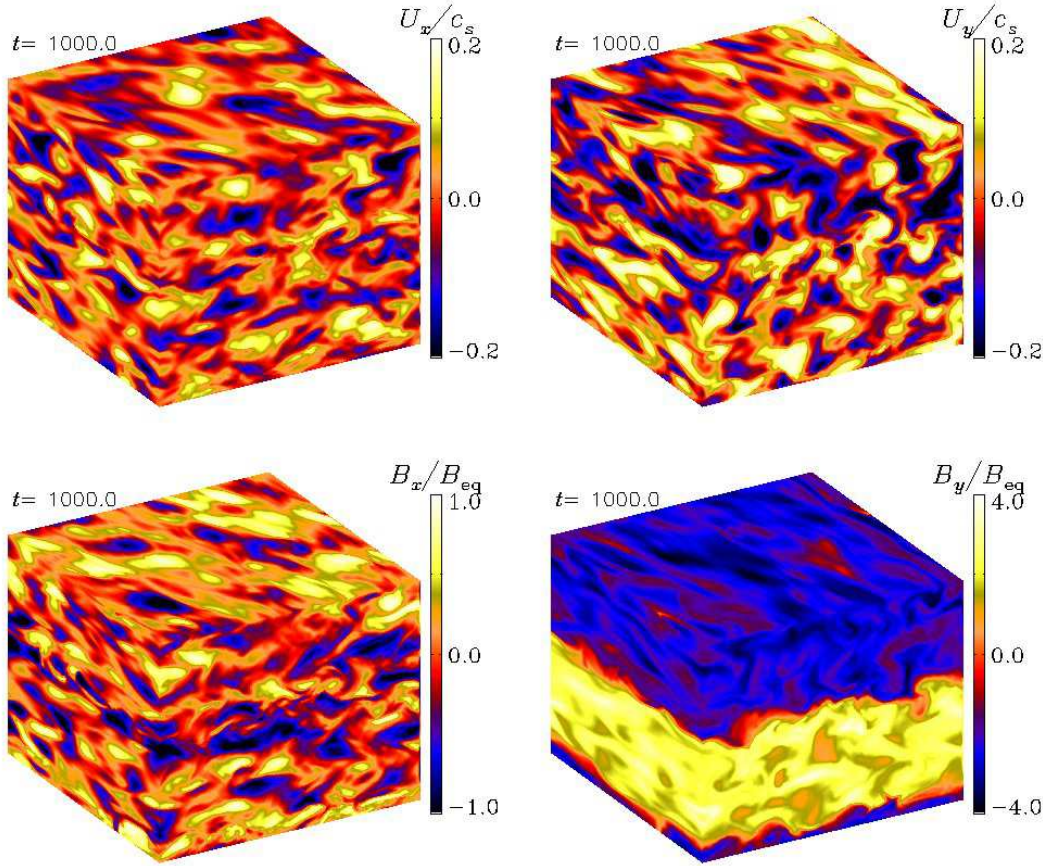}
\caption{$U_{\rm x}$, $U_{\rm y}$, $B_{\rm x}$, and $B_{\rm y}$
  at the periphery of the domain
  from a run with $\Rm\approx209$, $\Sh\approx-0.18$, and $\Pm=10$.
  Note the absence of any large-scale pattern in $U_y$, but a rather
  pronounced pattern in $B_y$ with a fairly abrupt change of sign.
}.
\label{boxes}
\end{figure*}

We use periodic boundary conditions in the $y$- and $z$-directions and
shearing-periodic boundary conditions in the $x$-direction.
The main control parameters in our simulations are the magnetic
Reynolds and Prandtl numbers, as well as the shear parameter,
\EQ
\Rm={u_{\rm rms}\over\eta k_{\rm f}},\quad
\Pm={\nu\over\eta},\quad
\Sh={S\over u_{\rm rms}\kf}.
\EN
By setting $k_1 = c_{\rm s} = \rho_0= \mu_0 = 1$,
we obtain dimensionless units of length, velocity, density, and
magnetic field as
\begin{eqnarray}
[x]=k_1^{-1}, \quad [u] = c_{\rm s}, \quad [\rho] = \rho_0, \quad [B] = \sqrt{\mu_0 \rho_0}c_{\rm s} .
\end{eqnarray}
We solve the governing equations using the {\sc Pencil Code}%
\footnote{\texttt{http://www.nordita.org/software/pencil-code}}
which is a high-order finite-difference code (sixth order in space
and third order in time) for solving partial differential equations
on massively parallel machines.

The three essential parameters varied in the present study are the
root-mean-square (rms) velocity (by changing the forcing amplitude $f_0$), the relative
helicity of the turbulence (by changing $\sigma$) and the microscopic magnetic
diffusivity $\eta$. The rms velocity varies between $0.03-0.2$,
corresponding to $f_0=0.01-0.05$. The large-scale shear
varies between $S=-0.05, \ldots, -0.2$ and leads to shear parameters
$\Sh\approx -0.1, \ldots, -1.4$. The high $\Sh$ cases studied in
Brandenburg \& K\"apyl\"a (2007) tend to produce very strong
magnetic fields. The purpose of the study is to investigate the
saturated state with different values of the mean magnetic field, 
$\meanv{B}/\Beq$, so we need
to regulate its value. One way to decrease the magnetic fields is to
increase the forcing amplitude to reduce the relative importance of the
$\Omega$ effect. On the other hand, varying the value of $\sigma$ can
be used to regulate the $\alpha$ effect with given $\Sh$ and $\Rm$.
This procedure was first adopted by Maron \& Blackman (2002) and later
by Brandenburg et al.\ (2002) in order to study the onset of large-scale
dynamo action.
In most cases $\sigma$ is varied between $0.05$ and 1, but small values
do not always lead to clear large-scale dynamo action. Performing a
large set of runs we have been able to cover a range of
$(\meanv{B}/\Beq)^2$ from $0.3$ to $270$ and the magnetic Reynolds
number varies between 1.5 and roughly 210.
The numerical resolution of the simulations varies between $64^3$ and
$256^3$ grid points.
A summary of the runs is given in Table~1.

\section{Formalism}

In this section we briefly review some of the main predictions
of the dynamical quenching model as applied to homogeneous shear
flows (BB02).
In this approach, the quenching of $\alpha$ comes from an additional
contribution that is proportional to the small-scale current helicity,
which builds up at the same time as the $\alpha$ effect produces
large-scale magnetic fields.
The sign of the small-scale current helicity is opposite to that
of the large-scale current helicity which, in turn, is also equal
to the sign of the $\alpha$ effect.
The evolution of the small-scale current helicity is calculated
from the magnetic helicity equation.
Using this equation together with the mean-field induction equation,
BB02 derived expressions for the saturation amplitude, the cycle
frequency, and the ratio of toroidal to poloidal field amplitudes
in closed form.
We begin with a more precise definition of $k_{\rm m}$ in terms of the
resulting mean magnetic field via
\EQ
k_{\rm m}^2=\mu_0\bra{\meanJJ\cdot\meanBB}/\bra{\meanAA\cdot\meanBB},
\EN
where $\mu_0$ is the vacuum permeability, overbars denote mean
quantities (later we shall specialize to horizontal averages), and
angular brackets denote volume averaging.
Preliminary results on turbulent diffusivity determined from the
cycle frequency were presented in Brandenburg \& K\"apyl\"a (2007).
In the present study, we explore a much wider range of parameters and
seek to understand the quenching behavior of $\etat$ as a function
of the magnetic field strength.

For future reference we define at this point an analogously defined
wavenumber of the fluctuating field via
\EQ
k_{\rm f}^2=\mu_0\bra{\jj\cdot\bb}/\bra{\aaa\cdot\bb}.
\EN
We also define effective wavenumbers of the fluctuating and mean fields via
\EQ
\epsilon_{\rm f}k_{\rm f}=\mu_0\bra{\jj\cdot\bb}/\bra{\bb^2},\label{equ:epsfkf}
\EN
\EQ
\epsilon_{\rm m}k_{\rm m}=\mu_0\bra{\meanJJ\cdot\meanBB}/\bra{\meanBB^2}, \label{equ:epsmkm}
\EN
where $\epsilon_{\rm f}$ and $\epsilon_{\rm m}$ are the fractional helicities 
of the fluctuating and mean fields, respectively.
We assume these two fractional helicities to be positive, but $k_{\rm f}$
and $k_{\rm m}$ can have either sign.
In the stationary state the total current helicity must vanish for a closed
system (Brandenburg 2001), which means that $k_{\rm f}$
and $k_{\rm m}$ must have opposite sign.
Throughout this work the forcing function has positive helicity,
so $k_{\rm f}>0$, and therefore $k_{\rm m}<0$.

BB02 compared their results with those of direct simulations of
Brandenburg et al.\ \cite{Brandenburgea2001}.
Theoretically, in the linear regime and at large magnetic Reynolds
numbers, $-k_{\rm m}$ can be as large as $k_{\rm f}/2$
(Brandenburg et al.\ 2002), but in the nonlinear regime,
$-k_{\rm m}$ will decrease until it reaches $k_1$
(Brandenburg 2001).
Here we consider nonlinear solutions and will therefore assume $-k_{\rm m}=k_1$.

BB02 predicted that the saturation level of the mean magnetic field is given by
\EQ
\frac{B_{\rm fin}^2}{\Beq^2} =
\frac{\epsilon_{\rm f}\kef}{\epsilon_{\rm m}k_1}
- (1 + \Rm^{-1})\;. \label{equ:b2pred}
\label{BfinBeq}
\EE
Here, $\epsilon_{\rm f}k_{\rm f}$ and $\epsilon_{\rm m}k_1$ are the
effective wavenumbers corresponding to the scale of the forcing and the
mean field, respectively, and are defined via Equations~(\ref{equ:epsfkf}) and (\ref{equ:epsmkm}). 
The equipartition value of the magnetic field is defined via
\EQ
\Beq=\bra{\mu_0 \rho \bm{u}^2}^{1/2}.
\EN
According to the calculations of BB02, $\epsm$ is directly proportional
to the ratio of cycle frequency to shear rate.
Furthermore, $\epsm$ is also proportional to the ratio of
poloidal to toroidal magnetic field amplitudes.
We have therefore multiple checks on the consistency of
this simple model.

%

\begin{deluxetable*}{cccccccccccc}
\tabletypesize{\scriptsize}
\tablecaption{Summary of the different sets of runs. Here, 
  $\tilde{B}_{\rm rms}=\brms/\Beq$, 
  $\brac{\tilde{\mean{B}}_x^2}^{1/2}=\brac{\mean{B}_x^2}^{1/2}/\Beq$, and
  $\brac{\tilde{\mean{B}}_y^2}^{1/2}=\brac{\mean{B}_y^2}^{1/2}/\Beq$.}
\tablewidth{0pt}
\tablehead{
\colhead{Set} & \colhead{grid} & \colhead{$\Rm$} & \colhead{$\Pm$} & \colhead{$\kef/k_1$} &
\colhead{$\Sh$} & \colhead{$\sigma$} & \colhead{$\tilde{B}_{\rm rms}$} &
\colhead{$\brac{\tilde{\mean{B}}_x^2}^{1/2}$} & \colhead{$\brac{\tilde{\mean{B}}_y^2}^{1/2}$}
}
\startdata
A   & $64^3$  & 1.5        & 1  & 10 & -0.72 & $0.2\ldots1$  & $8.0\ldots13.0$ & $0.14\ldots0.22$ & $7.2\ldots11.6$ \\ 
B   & $64^3$  & $46\ldots94$ & 10 &  5 & $-0.33\ldots-0.16$ & $0.05\ldots1$ & $1.5\ldots2.4$  & $0.05\ldots0.22$ & $0.95\ldots1.8$ \\ 
C   & $64^3$  & $27\ldots35$ & 10 &  5 & $-0.57\ldots-0.44$ & $0.05\ldots0.7$ & $2.0\ldots9.1$ & $0.05\ldots0.09$ & $1.5\ldots8.2$ \\ 
D   & $64^3$  & $11\ldots13$ & 10  & 5 & $-1.43\ldots-1.16$ & $0.05\ldots1$ & $2.7\ldots17.9$ & $0.03\ldots0.10$ & $1.8\ldots16.2$ \\ 
E   & $128^3$ & 11 & 10  & 5 & $-1.42\ldots-1.36$ & $0.2\ldots1$ & $7.9\ldots14.4$ & $0.07\ldots0.10$ & $7.2\ldots13.1$ \\ 
F   & $128^3$ & $131\ldots209$ & 10  & 5 & $-0.29\ldots-0.18$ & $0.05\ldots1$ & $1.2\ldots2.9$ & $0.04\ldots0.20$ & $0.5\ldots2.2$ \\ 
G   & $256^3$ & 29 & 25  & 5 & $-1.38\ldots-1.29$ & $0.2\ldots1$ & $8.3\ldots15.9$ & $0.05\ldots0.06$ & $7.5\ldots14.3$ \\ 
H   & $64^3\ldots128^3$ & $1.4\ldots181$ & $0.5\ldots50$  & 5 & $-0.13\ldots-0.10$ & 1 & $1.7\ldots3.1$ & $0.23\ldots0.55$ & $1.2\ldots2.3$ \\ 
\enddata
\end{deluxetable*}

\section{Results}
\label{Results}

\subsection{Saturation level of the magnetic field}
In most cases we find oscillatory solutions with dynamo waves
propagating in the positive $z$ direction.
This is indeed expected from mean-field theory, according to which
the direction of propagating is given by the sign of the product of
$\alpha$ effect and shear.
Indeed, positive helicity in the forcing should result in a negative
$\alpha$ effect which, together with negative shear, predicts a
direction of propagation in the positive $z$ direction.

\FFig{boxes} shows the streamwise components of velocity and magnetic
field from a run with $\Rm\approx209$, $\Sh\approx-0.18$, and $\Pm=10$.
The velocity field is irregular while the $y$-component of the magnetic field 
exhibits clear large-scale structure.
The $x$-component of the field also has a systematic large-scale component,
but it
is hard to see in a single snapshot and without horizontal averaging
because its amplitude is much lower than that of the $y$-component.

\begin{figure}[t]
\centering\includegraphics[width=\columnwidth]{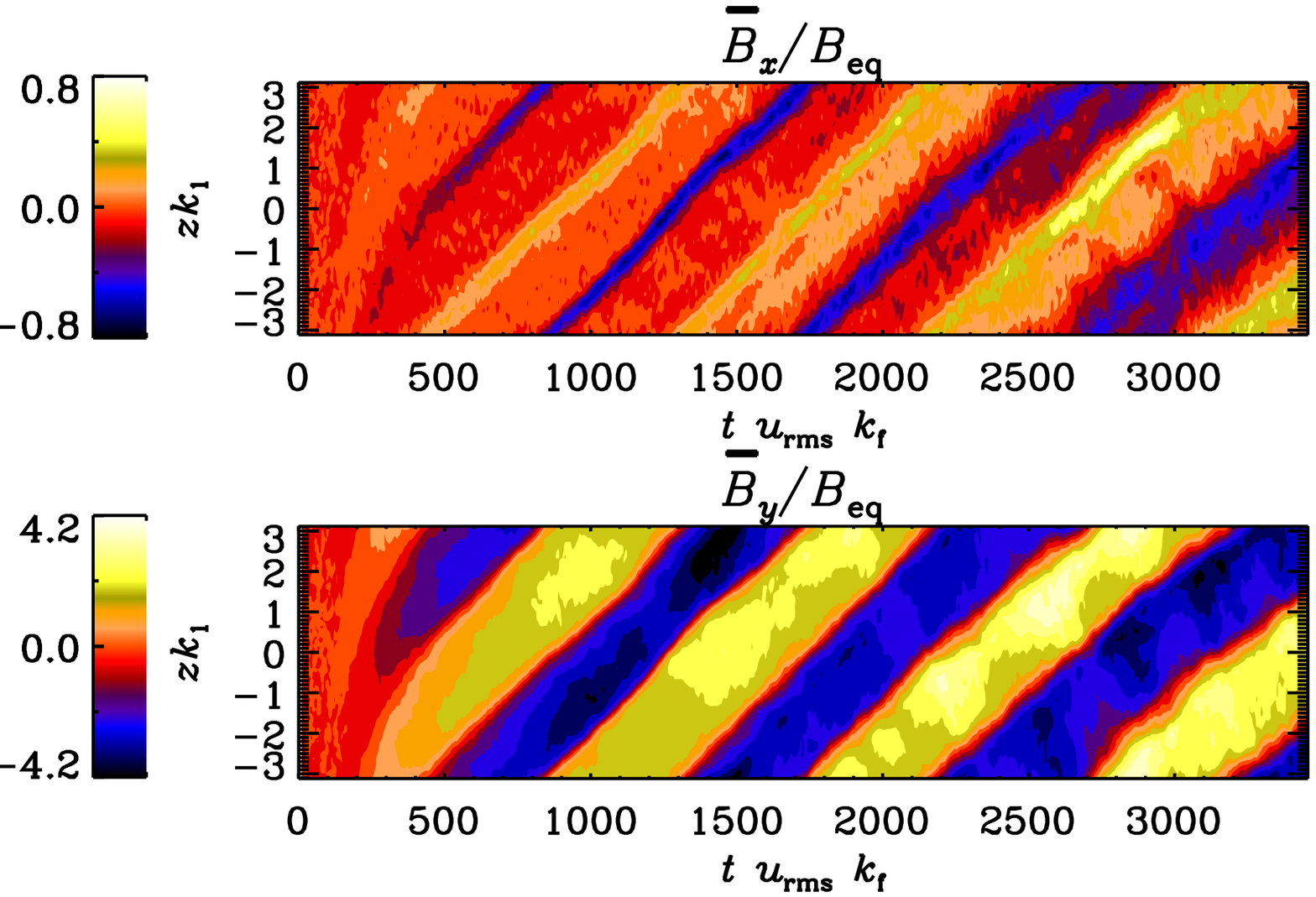}
\caption{Space-time diagrams of $\mean{B}_x(z,t)$ (upper panel) and
  $\mean{B}_y(z,t)$ (lower panel).
  From a run with ${\Rm}\approx 209$, $\Sh\approx-0.18$ and
  $\Pm=10$.}
   \label{fig:st}
\end{figure}

This dynamo wave is well seen in animations showing the $z$ dependence
of $B_y$ vs.\ time, but it becomes particularly clear when the field
is averaged over the horizontal directions, indicated
here by an overbar.
An example of $\meanB_x(z,t)$ and $\meanB_y(z,t)$ is shown in \Fig{fig:st}.
Note again the sharp sign changes of $\meanB_y(z,t)$.
One can now see that the locations of these sharp sign changes coincide with
the locations where $\meanB_x(z,t)$ achieves positive or negative extrema.
In fact, by comparing the two panels of \Fig{fig:st} one can verify
that a positive extremum of $\meanB_x(z,t)$ leads to a change of sign
of $\meanB_y(z,t')$ from a positive value at $t'<t$ to a negative value
at $t'>t$, and vice versa.
This is explained by the fact that shear is negative, i.e. $S<0$, which
turns a positive $\meanB_x$ into a negative $\meanB_y$.
 
\begin{figure}[t]
\centering\includegraphics[width=\columnwidth]{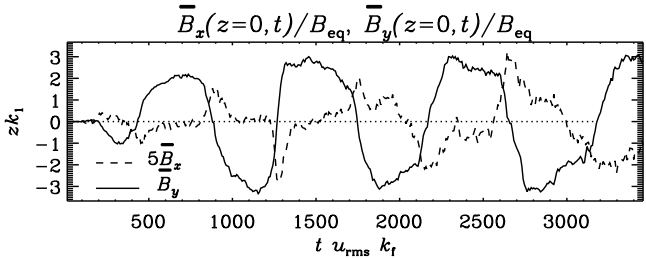}
\caption{Five times $\mean{B}_x$ (dashed line) and $\mean{B}_y$ (solid
  line) in the midplane, i.e.\ $z=0$. From the same run as in
  \Fig{fig:st}.}
   \label{fig:pbxby}
\end{figure}

The dynamo wave has a typical anharmonic shape (see \Fig{fig:pbxby}),
just as has been seen before both in mean-field models (Stix 1972)
as well as in direct simulations (Brandenburg et al.\ 2001).
\FFig{fig:pbxby} shows flat positive or negative plateaus in $\meanB_y$,
during which the $\meanB_x$ field was weak, but with a clear time
derivative: while $\meanB_y$ is positive, $\partial\meanB_x/\partial t$
is also positive.
A more careful look reveals that the plateaus are not completely flat,
but have a negative time derivative when $\meanB_y$ is positive
(and a positive time derivative when $\meanB_y$ is negative).
This relation can be interpreted as being due to a negative $\alpha$ effect
in the relation
\EQ
{\partial\meanBB\over\partial t}=\nab\times(\alpha\meanBB)+\ldots\,.
\EN
Using the fact that there is a dynamo wave with propagation speed $c$,
we have $\meanBB=\meanBB(z-ct)$.
Assuming constant $\alpha$, this means that $c\meanB_x'=\alpha\meanB_y'$.
Since $c$ is positive (compare \Fig{fig:st}), the opposite signs of the time
derivatives of $\meanB_x$ and $\meanB_y$ suggest that $\alpha<0$.
This is in agreement with the fact that the kinetic helicity is positive
and mean-field theory suggesting that for isotropic turbulence in the
high-conductivity limit (Moffatt 1978, Krause \& R\"adler 1980) $\alpha$
is a negative multiple of the kinetic helicity.

\begin{figure}[t]
\centering\includegraphics[width=\columnwidth]{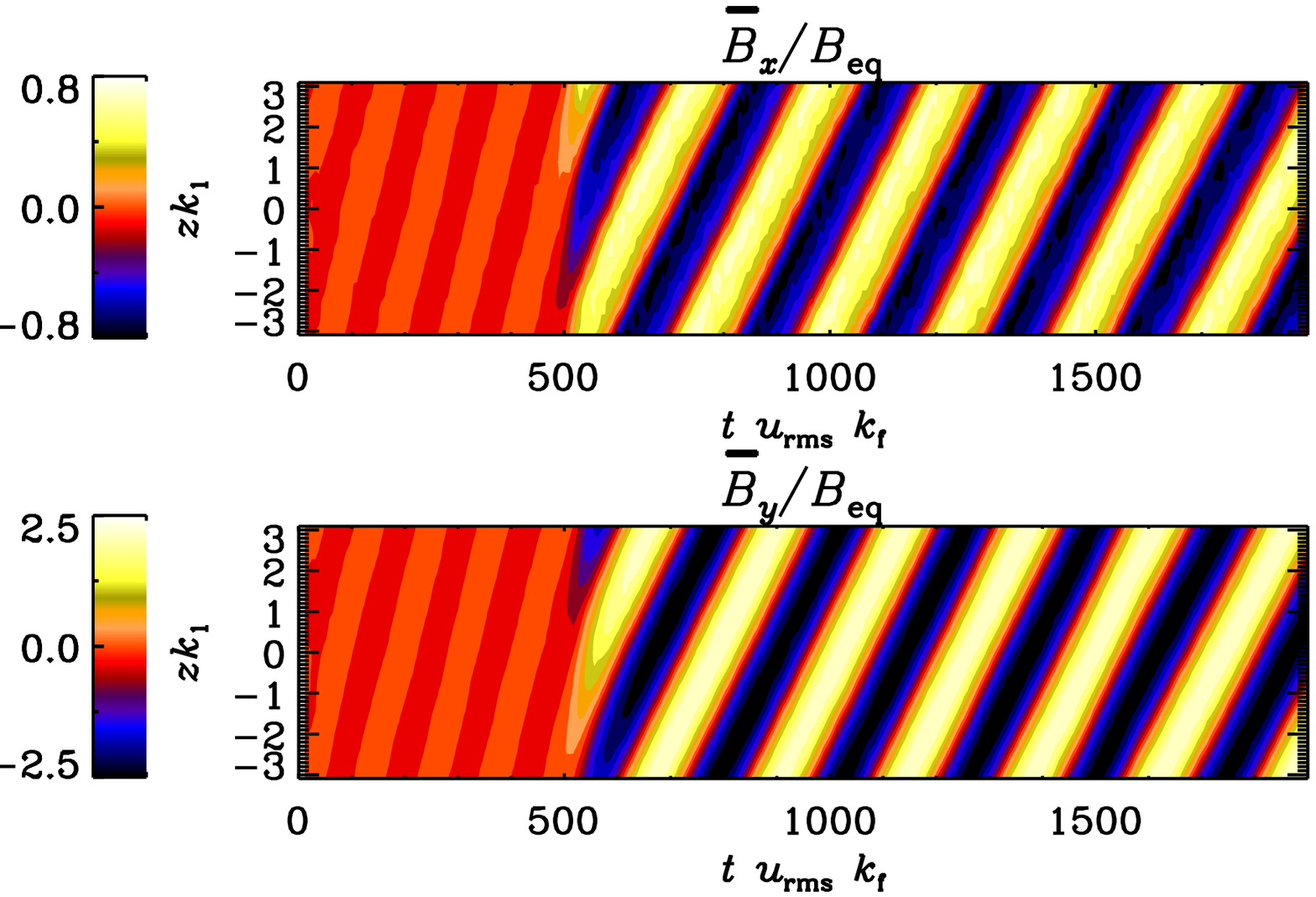}
\caption{Space-time diagrams of $\mean{B}_x(z,t)$ (upper panel) and
  $\mean{B}_y(z,t)$ (lower panel) from a run with ${\Rm}\approx 1.4$,
  $\Sh\approx-0.13$ and $\Pm=1$.}
   \label{fig:st_h64b_per5}
\end{figure}

\begin{figure}[t]
\centering\includegraphics[width=\columnwidth]{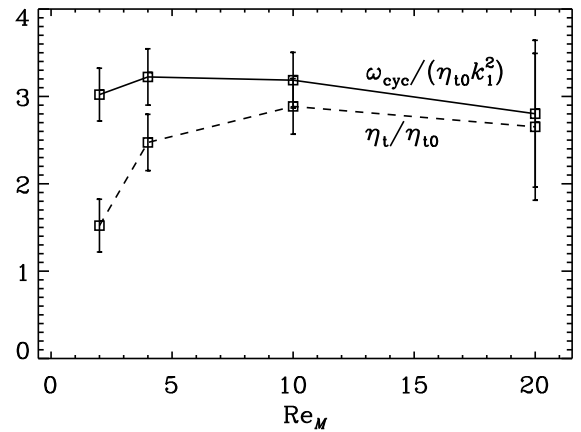}
\caption{Cycle frequency (solid line) and turbulent diffusivity from Equation~(\ref{ocyc}) (dashed line) as functions of $\Rm$ in the kinematic regime.}
   \label{fig:pocyckin}
\end{figure}

The oscillations are also discernible in the kinematic regime
especially in runs where $\Rm$ is small enough, so that the small-scale
dynamo is not excited (see \Fig{fig:st_h64b_per5}).
The cycle frequency in the kinematic regime appears to be roughly constant in
the range $\Rm=2\ldots 10$.
In the following, however, we will be concerned with the nonlinear regime
where the dynamo has reached saturation.
We study the dependence of the value of $\omega_{\rm cyc}$ on $\Rm$.
Using Equation~(\ref{ocyc}) we then calculate $\etaT$ and hence $\etat$.
\FFig{fig:pocyckin} shows that the resulting value of $\etat$ has
a value roughly $2.5\etatz$ for $\Rm\ga4$, where
\EQ
\etatz = \onethird \tau \mean{\bm{u}^2}\;,
\label{etatz}
\EN
and $\tau$ is the correlation time of the turbulence. The subscript zero 
refers to the kinematic case which is valid when the magnetic field is weak. 
For small $\Rm$ the correlation time is no longer determined by the
turbulence but rather the microscopic diffusivity $\eta$. Thus, in that
limit $\etat$ is expected to decreases proportionally to $\eta$. The
markedly lower value of $\etat$ for $\Rm\approx2$ can be interpreted
in terms of this behavior.

\begin{figure}[t]
\centering\includegraphics[width=\columnwidth]{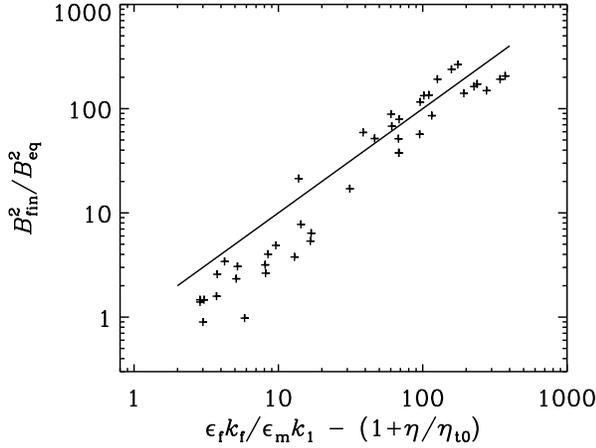}
\caption{$B_{\rm fin}^2/\Beq^2$ as a function of
$\epsf\kf/\epsm k_1-(1+\eta/\etatz)$.
The diagonal is shown for comparison.
Note the reasonable agreement with theory (solid line).
}\label{pepsf}
\end{figure}

For $\Rm \gg 1$ the saturation formula given in
Equation~(\ref{equ:b2pred}) predicts that the ratio $B_{\rm fin}^2/\Beq^2$
should be linearly proportional to $\epsf\kf/\epsm k_1-(1+\eta/\etatz)$.
This is indeed in reasonable agreement with the numerical data;
see Figure~\ref{pepsf}.

\begin{figure}[t]
\centering\includegraphics[width=\columnwidth]{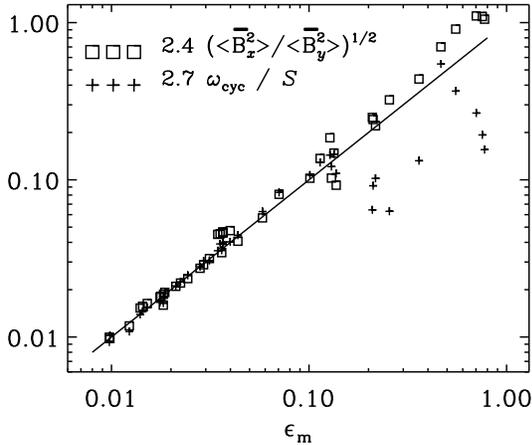}
\caption{
Scatter plot of poloidal to toroidal field ratio (squares,
scaled by factor 2.7) and normalized cycle frequency (plus signs,
scaled by factor 2.4) vs.\ $\epsilon_{\rm m}$.
Note that for $\epsilon_{\rm m}<0.1$ both squares and plus signs
scatter tightly around the diagonal.
}\label{peps_all}
\end{figure}

\begin{figure}[t]
\centering\includegraphics[width=\columnwidth]{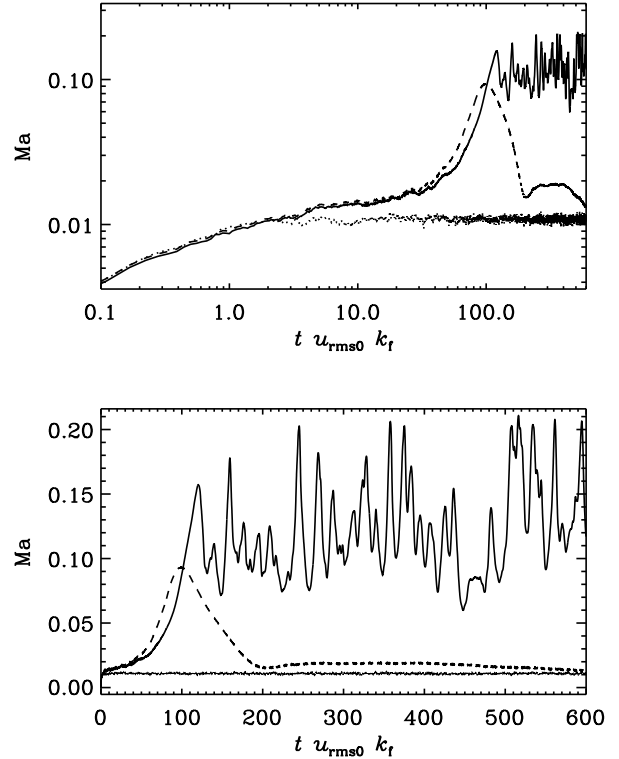}
\caption{
  Root mean square Mach number for two shearing runs with the same
  initial parameters ($S=-0.2$, $\nu=5\cdot10^{-3}$) with (dashed
  line) and without (solid line) magnetic fields.
  The dotted line shows the result for a non-shearing hydrodynamic
  run for comparison.
}
\label{pvdyn}
\end{figure}

\begin{figure}[t]
\centering\includegraphics[width=\columnwidth]{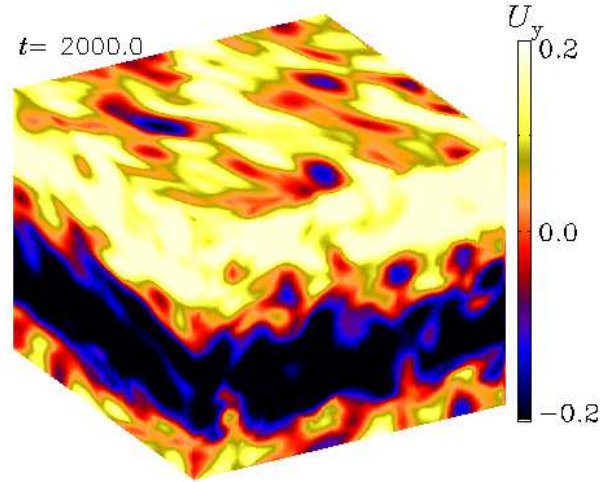}
\caption{$U_{\rm y}$ at the periphery of the domain
  from a run with $\Rm\approx80$, $\Sh\approx-0.19$, and $\Pm=10$,
  at a resolution of $64^3$ mesh points at a time when the magnetic field
  is saturated and a large-scale vorticity dynamo develops.}.
\label{64c4_Uy}
\end{figure}

BB02 gave two independent formulae for $\epsilon_{\rm m}$, one in terms of the
resulting ratio of poloidal to toroidal mean fields, $Q^{-1}$, where
$Q^2=\brac{\mean{B}_y^2}/\brac{\mean{B}_x^2}$,
\EQ
\epsilon_{\rm m}=\sqrt{2}Q^{-1}
\equiv\sqrt{2}\,(\bra{\meanB_x^2}/\bra{\meanB_y^2})^{1/2},
\label{equ:epsmkm2}
\EN
and one in terms of the resulting ratio of the cycle frequency to the
shear frequency,
\EQ
\epsilon_{\rm m}=2\omega_{\rm cyc}/S. \label{equ:epsmkm3}
\EN
Furthermore, assuming $k_{\rm m}=-k_1$ in Equation~(\ref{equ:epsmkm}), we 
arrive at
\EQ
\epsilon_{\rm m}=-\mu_0\frac{\langle\meanv{J}\cdot\meanv{B}\rangle}{k_1 \langle \meanv{B}^2 \rangle}. \label{equ:epsmkm1}
\EN
We use the definition of $\epsilon_{\rm m}$ given in
Equation~(\ref{equ:epsmkm1}) as the benchmark value to compare the other
two expressions to. A scatter plot of the results is given in
Figure~\ref{peps_all}. We find that the values given by
Equations~(\ref{equ:epsmkm2}) and (\ref{equ:epsmkm3}) are consistently
below the value of $\epsm$ as given by Equation~(\ref{equ:epsmkm1}),
but that the trend as a function of $\sigma$ is
the same for all three quantities. Thus, we apply scaling factors of
the order of unity for the expressions in Eqs.~(\ref{equ:epsmkm2}) and
(\ref{equ:epsmkm3}) in Figure~\ref{peps_all} so that the data fall onto
the same line.
An exception are a few points for the cycle frequency, given by
Equation~(\ref{equ:epsmkm3}), where there are departures.
The points deviating from the trend all belong to runs that have the weakest
mean fields and occur for $\epsilon_{\rm m}>0.1$.

\subsection{Effects from the vorticity dynamo}

A few runs exhibit signs of intermittent or continuous large-scale
vorticity generation in the saturated
state of the dynamo, reminiscent of a vorticity dynamo 
(Elperin et al.\ 2003; Yousef et al.\ 2008a,b).
In Figure~\ref{pvdyn}, we compare the evolution of the rms Mach number of
the total (mean and fluctuating) velocity both for hydrodynamic runs
with and without shear, as well as a run with magnetic fields and shear.
It is clear that in the absence of shear the rms Mach number reaches
its final level quite quickly.
During the first few tens of turnover times the same level is also maintained
in the presence of shear, but the velocity gradually increases
to much larger values and reaches a new saturation level that is
approximately 10 times larger.
During that time a large-scale velocity field develops throughout the
entire domain; see \Fig{64c4_Uy} for a typical example of this behavior.
In the hydromagnetic case the vorticity dynamo is quenched when the magnetic
field grows to high enough level but in cases of weaker magnetic
fields this quenching is often only partial as demonstrated by the
dashed line in Figure~\ref{pvdyn}.
Unlike the anisotropic kinetic $\alpha$ effect (Frisch et al.\ 1987), which
is also suppressed by magnetic fields (Brandenburg \& von Rekowski 2001),
and works only for $\Rey\equiv\Rm/\Pm\leq8$, the vorticity dynamo has been
shown to work well up to $\Rey\approx100$.
A more thorough study of the vorticity dynamo can be found elsewhere
(K\"apyl\"a et al.\ 2009b).

The overall behavior of such runs is quite different from those where
a large-scale flow did not develop.
We have therefore excluded these runs from the plots for the sake of clarity.
All in all the agreement between the three independent checks of
$\epsilon_{\rm m}$ is remarkably good especially for low values 
of $\epsilon_{\rm m}$.

\subsection{Turbulent diffusivity quenching}

According to the first order smoothing approximation
(Moffatt 1978, Krause \& R\"adler 1980),
the turbulent diffusivity for isotropic turbulence
in the high-conductivity limit is given by Equation~(\ref{etatz}).
Simulations of forced turbulence without shear suggest that, independent
of the value of the forcing wavenumber and amplitude, the Strouhal number
is around unity (Brandenburg \& Subramanian 2005a, 2007), i.e.\
\EQ
\St \equiv \kef \urms \tau \approx 1\;.
\EN
In the kinematic regime,
the magnetic diffusivity can therefore be written as
\EQ
\etatz = \onethird \St\ \urms \kef^{-1} \approx \onethird \urms \kef^{-1}\;. \label{equ:etatz}
\EN
Recent numerical simulations employing the test-field procedure have
shown that in the kinematic case, Equation~(\ref{equ:etatz}) is confirmed
for $\Rm$ between 1 and 200; see Sur et al.\ \cite{Surea2008}.
In order
to study the quenching of $\etat$, i.e.\ the dependence on $\meanBB$,
we normalize our results with the value of $\etatz$ from a
simulation without magnetic fields. 

Our setup is similar to that used in a number of related simulations
by Brandenburg et al.\ \cite{Brandea2008} with non-helical turbulence,
and Mitra et al.\ \cite{Mitraea2009} with helicity.
In those simulations the main focus was the determination of
turbulent transport coefficients in the linear regime.
In these studies it was found that the turbulent diffusivity increases
by a factor of a few as $|\Sh|$ increases from $0.1$ to unity.
More recently, this work has been extended to the nonlinear regime,
but so far only in the absence of shear (Brandenburg et al.\ 2008b).
However, there is then a potential difficulty in that there could be
additional terms in the functional form of the quenched $\alpha$
and $\etat$ tensors that could mimic turbulent diffusion, so the
split into two $\alpha$ and $\etat$ coefficients is not unique.

In the saturated state the growth rate of the large-scale field is zero.
Thus, this situation corresponds to the marginally
excited state where the cycle frequency is given by Equation~(1).
Measuring therefore the dynamo frequency in the saturated state gives the
(quenched) value of the turbulent diffusivity. 

\begin{figure}[t]
\includegraphics[width=\columnwidth]{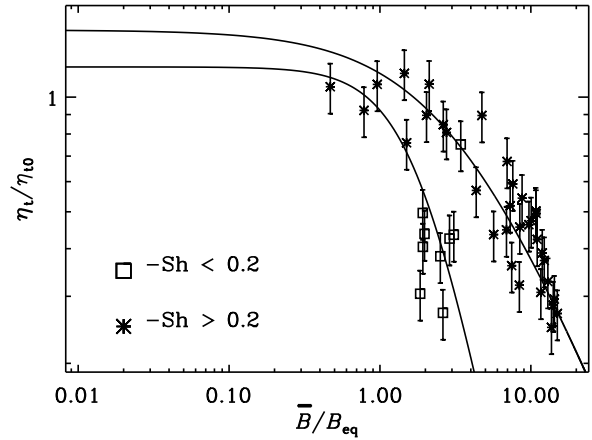}
\caption{
Turbulent diffusivity as a function of $|\meanv{B}|/\Beq$.
The squares and stars denote cases where $-\Sh < 0.2$ (weak shear)
and $-\Sh > 0.2$ (strong shear).
The upper curve is for $n=1$ and $\tilde\etatz/\etatz=1.5$,
while the second curve is for $n=2$ and $\tilde\etatz/\etatz=1.2$,
and $g=0.3$ in both cases.
}\label{fig:pall_etat}
\end{figure}

Results for all of the runs are shown
in Figure~\ref{fig:pall_etat}. The data scatter around the curve
\EQ
\etat={\tilde\etatz\over1+g(|\meanv{B}|/\Beq)^n},
\EN
with $\tilde\etatz=(1.2$--$1.5)\,\etatz$, $g=0.3$ and $n=1$--$2$
being fit parameters.
A factor greater than one in the definition of $\tilde\etatz$ reflects
the fact that in the kinematic regime, the turbulent magnetic diffusivity is enhanced
in the presence of shear (Brandenburg et al.\ 2008a, Mitra et al.\ 2009).
Note that the data for many of the runs with different magnetic Reynolds numbers
($\Rm=1\ldots210$) seem to fall roughly on the same line.
An asymptotic quenching of $\etat$ inversely proportional to $\meanBB$
instead of $\meanBB^2$
has been predicted analytically by Kitchatinov et al.\
(1994) and Rogachevskii \& Kleeorin (2001), and BB02 found $g\approx3$ to be
a good fit to the numerical simulations of Brandenburg et al.\
\cite{Brandenburgea2001}.
The fact that $g$ is here different suggests that it is perhaps not a
universal constant, but that it may depend on other parameters.

The amount of scatter in Figure~\ref{fig:pall_etat} does not depend
in any systematic way on the value of $\Sh$,
except for the lowest values of $\Sh$ for which the values of $\etat$
fall below the general trend. 
This, and the fact that $\etat$ seems to reach values above unity for
low magnetic field strengths, might be a consequence of the turbulent 
diffusivity being enhanced when shear increases (Brandenburg et al.\ 2008a,
Mitra et al.\ 2009).
In the present case the unquenched value of $\urms$ from a
non-shearing run with the same forcing amplitude has been used in the
definition of $\etatz$. If the $\urms$ from a hydrodynamic run with
shear is used, the excitation of the vorticity dynamo would enhance
its value by a factor of up to 5 (see Figure~\ref{pvdyn}). 

\begin{figure}[t]
\centering
\includegraphics[width=\columnwidth]{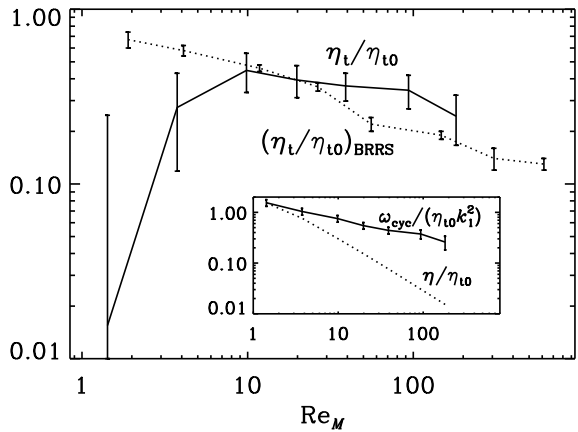}
\caption{Turbulent magnetic diffusivity as a function of $\Rm$, as
  obtained from the cycle frequency (solid line) for $\Sh=-0.1$ and $\sigma=1$, which translates to $\epsf\approx1$,
  compared with the corresponding result by
  Brandenburg et al.\ (2008b) using the test-field
  method ($\Sh=0$).
  The inset shows the $\Rm$ dependence of the cycle frequency
  $\omega_{\rm cyc}$ (expected to be proportional to $\etaT$;
  solid line) and the microscopic magnetic diffusivity (dotted line).}
\label{fig:peta_rm}
\end{figure}

In order to study the behavior of $\etat$ as a function of $\Rm$ a
set of runs were performed where $\Pm$ was varied between 0.5 and
50, keeping all the other parameters fixed, i.e.\ $\Sh\approx-0.1$ and
${\rm Re}=\urms/(\nu\kef)\approx4$. The results are shown in
Figure~\ref{fig:peta_rm}, which demonstrates that $\etat$, as inferred
from $\omega_{\rm cyc}$ using Equation~(\ref{ocyc}), decreases
by a factor of about 2 when $\Rm$ is increased from 10 to 180, while
$|\meanBB|/\Beq$ varies from 1.7 to 2.5, with the larger values occurring
at larger values of $\Rm$. 
The low value of $\etat$ for the smallest value of $\Rm$ might derive
from the fact that the simple model used to estimate the value of the
turbulent diffusivity may not be representative of this regime. On the
other hand, there appears to be a declining trend of $\etat$ as a
function of $\Rm$ approximately proportional to $\Rm^{1/3}$. However,
higher $\Rm$ simulations are needed to substantiate this.
On the other hand, data from nonlinear test-field calculations
by Brandenburg et al.\ (2008b) show that $\etat$ decreases by a factor of 5
when $\Rm$ is increased from 2 to 600.
Within error bars this is compatible with the present results.

\section{Conclusions}
\label{Conclusions}
Three dimensional direct numerical simulations of helically forced
turbulence with imposed large-scale shear have been used to study the
saturation level of the dynamo, its cycle frequency, and thereby
the turbulent diffusivity and its quenching with the magnetic field.
The parameters of the study were chosen such that a cyclic
large-scale magnetic field develops with the dynamo wave propagating
to the positive $z$-direction.

We find that the saturation level of the energy of the mean magnetic field
is compatible with a scaling of a quantity related to the ratio
of the fractional helicities of the small and large-scale fields, in accordance with
Equation~(\ref{BfinBeq}) and the prediction from nonlinear mean-field theory derived by BB02.
Furthermore, the three independent measures of $\epsilon_{\rm m}$ are
in reasonably good agreement, especially for small values of
$\epsilon_{\rm m}$.

The turbulent diffusivity is found to be quenched approximately
inversely proportional to the large-scale magnetic field strength.
This is in agreement with earlier analytical studies (Kitchatinov et al.\ 1994;
Rogachevskii \& Kleeorin 2001).
However, a small group of data points can also be fitted to a
quadratic dependence on the inverse field strength.
The dependence on $\Rm$ is found to be
weak which is also in accordance with the analytical studies. More
recent nonlinear test-field calculations (Brandenburg et al.\ 2008b),
indicate a similar dependence.

The present results may have implications for solar dynamo simulations.
Clearly, the hope is that cyclic reversals occur on time scales that
are asymptotically independent of the resistive time scale.
What is less clear, however, is when asymptotic behavior sets in.
Evidently, even for the highest $\Rm$ simulations presented here the
cycle frequency shows still a shallow decline with increasing magnetic
Reynolds number; see the inset of Figure~\ref{fig:peta_rm}.
On the other hand, the corresponding change in the turbulent magnetic
diffusivity is only by a factor of a few, even when the magnetic
Reynolds number changes by two orders of magnitude.
This suggests that higher resolution simulations are needed to have any
hope in seeing truly asymptotic behavior at large magnetic Reynolds numbers.

A modest level of quenching of $\etat$ would certainly still be compatible
with solar dynamo models.
In fact, already since the early 1970s it was clear that standard solar
dynamo models of $\alpha\Omega$ type show cycle frequencies that are too
high for realistic parameters, so the models predicted cycle periods of
about byears instead of 22 years (K\"ohler 1970).
A modest quenching of $\etat$ would therefore point in the right
direction, but one would hope that the decrease in $\etat$ would
eventually level off.
Unfortunately, this is not evident from any of the simulations
presented so far.

As we have argued in the introduction, we do not expect the results
for $\etat(\Rm)$ to depend on the presence or absence of magnetic
helicity fluxes.
However, this expectation should be verified using simulations.
Allowing for such fluxes would mean that we have to abandon the
assumption of periodic boundary conditions.
Although open boundary conditions would imply saturation on a dynamical
time scale (not the resistive one, as in the present case of periodic
boundaries) the saturation amplitude would scale inversely with $\Rm$,
if it was not for the shear-mediated helicity fluxes that allow for
large saturation amplitudes on a dynamical time scale (Brandenburg 2005,
K\"apyl\"a et al.\ 2008).

\acknowledgements
The authors acknowledge the hospitality of NORDITA during the program
``Turbulence and Dynamos.''  P.\ J.\ K.\ acknowledges financial support
from the Finnish Academy grant No.\ 112431 and A.\ B.\ acknowledges
support from the Swedish Research Council under grant No.\ 621-2007-4064.
The numerical simulations
were performed on the supercomputers hosted by CSC -- IT Center for
Science Ltd.\ in Espoo, Finland, who are financed by the Finnish
ministry of education.

\newcommand{\yana}[3]{ #1, {A\&A,} {#2}, #3}
\newcommand{\yapj}[3]{ #1, {ApJ,} {#2}, #3}
\newcommand{\yprl}[3]{ #1, {Phys.\ Rev.\ Lett.,} {#2}, #3}
\newcommand{\yan}[3]{ #1, {Astron.\ Nachr.,} {#2}, #3}
\newcommand{\ypp}[3]{ #1, {Phys.\ Plasmas,} {#2}, #3}
\newcommand{\ybook}[3]{ #1, {#2} (#3)}
{}


\end{document}